\documentclass[pdftex,twocolumn,epjc3]{svjour3}          

\RequirePackage[T1]{fontenc}

\smartqed  

\RequirePackage{graphicx}
\RequirePackage{mathptmx}      
\RequirePackage{flushend}
\RequirePackage[numbers,sort&compress]{natbib}
\RequirePackage[colorlinks,citecolor=blue,urlcolor=blue,linkcolor=blue]{hyperref}
\usepackage{upgreek}
\usepackage{siunitx}
\usepackage{lineno}
\usepackage{booktabs}
\usepackage{amsmath}
\usepackage{natbib,hyperref}
\usepackage{doi}

\newcommand{\ntwothree}{\text{N}_{23}\text{-}32}
\newcommand{\kr}{^{83\text{m}}\text{Kr}}

\journalname{Eur. Phys. J. C}

\begin{document}

\title{Measurement of the inhomogeneity of the KATRIN tritium source electric potential by high-resolution spectroscopy of conversion electrons from $\mathbf{^{83m}}$Kr}

\thankstext{co}{Corresponding authors: \mbox{m.boettcher@uni-muenster.de}, \mbox{caroline.fengler@kit.edu}, \mbox{moritz.machatschek@web.de}, \mbox{magnus.schloesser@kit.edu}}

\date{Received: date / Accepted: date}

\institute{
Department of Physics and Astronomy, University of North Carolina, Chapel Hill, NC 27599, USA\label{unc}\and
Institute for Astroparticle Physics~(IAP), Karlsruhe Institute of Technology~(KIT), Hermann-von-Helmholtz-Platz 1, 76344 Eggenstein-Leopoldshafen, Germany\label{iap}\and
Institute for Data Processing and Electronics~(IPE), Karlsruhe Institute of Technology~(KIT), Hermann-von-Helmholtz-Platz 1, 76344 Eggenstein-Leopoldshafen, Germany\label{ipe}\and
Institute for Nuclear Physics, University of M\"{u}nster, Wilhelm-Klemm-Str.~9, 48149 M\"{u}nster, Germany\label{muenster}\and
Istituto Nazionale di Fisica Nucleare (INFN) -- Sezione di Milano-Bicocca, Piazza della Scienza 3, 20126 Milano, Italy\label{infnbicocca}\and
Politecnico di Milano, Dipartimento di Elettronica, Informazione e Bioingegneria, Piazza L. da Vinci 32, 20133 Milano, Italy\label{polimi}\and
Istituto Nazionale di Fisica Nucleare (INFN) -- Sezione di Milano, Via Celoria 16, 20133 Milano, Italy\label{infnmilano}\and
Department of Physics, Faculty of Science, Chulalongkorn University, Bangkok 10330, Thailand\label{chulalongkorn}\and
Department of Physics, Carnegie Mellon University, Pittsburgh, PA 15213, USA\label{cmu}\and
Nuclear Physics Institute (NPI),  Czech Academy of Sciences, 25068 \v{R}e\v{z}, Czech Republic\label{npi}\and
Institute of Experimental Particle Physics~(ETP), Karlsruhe Institute of Technology~(KIT), Wolfgang-Gaede-Str.~1, 76131 Karlsruhe, Germany\label{etp}\and
Technical University of Munich, TUM School of Natural Sciences, Physics Department, James-Franck-Stra\ss e 1, 85748 Garching, Germany\label{tum}\and
Department of Physics, Faculty of Mathematics and Natural Sciences, University of Wuppertal, Gau{\ss}str.~20, 42119 Wuppertal, Germany\label{wuppertal}\and
Center for Experimental Nuclear Physics and Astrophysics, and Dept.~of Physics, University of Washington, Seattle, WA 98195, USA\label{washington}\and
Laboratory for Nuclear Science, Massachusetts Institute of Technology, 77 Massachusetts Ave, Cambridge, MA 02139, USA\label{massit}\and
Dipartimento di Fisica, Universit\`{a} di Milano - Bicocca, Piazza della Scienza 3, 20126 Milano, Italy\label{umilano}\and
Triangle Universities Nuclear Laboratory, Durham, NC 27708, USA\label{tunl}\and
Max Planck Institute for Physics, Boltzmannstr. 8, 85748 Garching, Germany\label{mpp}\and
School of Physics and Center of Excellence in High Energy Physics and Astrophysics, Suranaree University of Technology, Nakhon Ratchasima 30000, Thailand\label{suranaree}\and
Max-Planck-Institut f\"{u}r Kernphysik, Saupfercheckweg 1, 69117 Heidelberg, Germany\label{mpik}\and
IRFU (DPhP \& APC), CEA, Universit\'{e} Paris-Saclay, 91191 Gif-sur-Yvette, France\label{saclay}\and
Nuclear Science Division, Lawrence Berkeley National Laboratory, Berkeley, CA 94720, USA\label{lbnl}\and
Institute for Technical Physics~(ITEP), Karlsruhe Institute of Technology~(KIT), Hermann-von-Helmholtz-Platz 1, 76344 Eggenstein-Leopoldshafen, Germany\label{itep}\and
Departamento de Qu\'{i}mica F\'{i}sica Aplicada, Universidad Autonoma de Madrid, Campus de Cantoblanco, 28049 Madrid, Spain\label{madrid}\and
Institut f\"{u}r Physik, Humboldt-Universit\"{a}t zu Berlin, Newtonstr.~~15, 12489 Berlin, Germany\label{berlin}\and
Institute for Theoretical Astrophysics, University of Heidelberg, Albert-Ueberle-Str.~2, 69120 Heidelberg, Germany\label{uhd}\and
Institut f\"{u}r Physik, Johannes-Gutenberg-Universit\"{a}t Mainz, 55099 Mainz, Germany\label{mainz}\and
Also affiliated with Department of Physics, Duke University, Durham, NC, 27708, USA\label{duke}\and
Also affiliated with Department of Physics, Washington and Jefferson College, Washington, PA 15301, USA\label{wnj}\and
Also affiliated with Oak Ridge National Laboratory, Oak Ridge, TN 37831, USA\label{ornl}
}

\author{KATRIN Collaboration\\\\ 
H.~Acharya\thanksref{unc}\and
M.~Aker\thanksref{iap}\and
D.~Batzler\thanksref{iap}\and
A.~Beglarian\thanksref{ipe}\and
J.~Beisenk\"{o}tter\thanksref{muenster}\and
M.~Biassoni\thanksref{infnbicocca}\and
B.~Bieringer\thanksref{muenster}\and
Y.~Biondi\thanksref{iap}\and
F.~Block\thanksref{iap}\and
B.~Bornschein\thanksref{iap}
% ...
}

\maketitle

\begin{abstract}
Precision spectroscopy of the electron spectrum of the tritium $\upbeta$-decay near the kinematic endpoint is a direct method to determine the effective electron antineutrino mass. The KArlsruhe TRItium Neutrino (KATRIN) experiment aims to determine this quantity with a sensitivity of better than $\SI{0.3}{\electronvolt}$ ($90\,\%$~C.L.). An inhomogeneous electric potential in the tritium source of KATRIN can lead to distortions of the $\upbeta$-spectrum, which directly impact the neutrino-mass observable. This effect can be quantified through precision spectroscopy of the conversion-electrons of co-cir\-cu\-la\-ted metastable $\kr$.
Therefore, dedicated, several-weeks long measurement campaigns have been performed within the KATRIN data taking schedule. 
In this work, we infer the tritium source potential observables from these measurements, and present their implications for the neutrino-mass determination.
\end{abstract}

\section{Introduction} 
The discovery of neutrino flavor oscillations \cite{Fukuda1998, Ahmad2002} confirmed that neutrinos possess mass, contrary to the original assumption in the Standard Model of particle physics. Although oscillation experiments provide insights into the differences in squared neutrino-mass eigenvalues, the absolute scale of neutrino masses remains a key unknown parameter in cosmology, nuclear, particle, and astroparticle physics. 

Precision measurements of the kinematics of weak decays, notably tritium $\upbeta$-decay, are the most model-indepen\-dent experimental approaches for determining the neutrino mass \cite{Formaggio2021}.
All recent $\upbeta$-decay experiments have employed the molecular form T$_2$ as electron source \cite{Robertson1991, Stoeffl1995, Kraus2005, Aseev2011}, which follows the decay scheme: 
\begin{equation}
    \mathrm{T_2} \to\, ^3\mathrm{HeT}^+ +\mathrm{e}^{-} + \bar{\nu}_\mathrm{e}.
\end{equation}
The decay energy is distributed to the kinetic energy and rest mass of the generated particles, as well as to the internal excitations of the $^3\mathrm{HeT}^+$ molecular ion. Analysis of the electron energy near the endpoint of the $\upbeta$ spectrum allows the determination of the effective squared neutrino mass $m_\upbeta^2=\sum_i |U_{\mathrm{e}i}|^2\,m_i^2$.

The KArlsruhe TRItium Neutrino experiment (KATRIN) \cite{Design2021, Lokhov2022} uses a gaseous molecular tritium source with a high activity of $A=10^{11}\,\si{\becquerel}$ together with a MAC-E filter type \cite{Beamson1980} integrating spectrometer with $\mathcal{O}(1\,\mathrm{eV})$ resolution. Recently, KATRIN obtained a new world-leading neutrino-mass upper limit of \SI{0.45}{eV} (\SI{90}{\percent} confidence level (C.L.)) from the first five measurement campaigns \cite{KMN1-5-paper}. KATRIN will continue to collect data until the end of 2025 and will achieve a final sensitivity of better than $\SI{0.3}{\electronvolt}$. 

The sensitivity of the neutrino-mass measurement depends not only on the statistics in the endpoint region, but also on the control, mitigation and understanding of systematic effects. 
In this paper we focus on the source section, in which the $\upbeta$-electrons are generated from tritium, and its systematic effects on the neutrino-mass observable.

 Spatial or temporal inhomogeneities of the source electric potential could lead to a distortion of the $\upbeta$-spectrum and thus to a bias of the neutrino-mass observable \cite{Belesev2007InvestigationExperiment}. The quantification of such inhomogeneities is performed by using gaseous $\kr$ as a calibration source in co-circulation with tritium and analyzing the $\kr$ conversion-electron line spectrum.

The mathematical framework for obtaining source-po\-ten\-tial observables from the $\kr$ spectra is described in \cite{Machatschek2021}. In the paper at hand, we present the results for source-potential observables obtained in two dedicated measurement campaigns. First, we introduce the KATRIN experiment and its tritium source in section~\ref{sec:KATRIN}, and the krypton measurement campaigns in section~\ref{sec:Kr_meas_campaign}. Thereafter, the data analysis is presented (section~\ref{sec:Ana}), the results are discussed, and the implications for the neutrino-mass analysis are laid out (section~\ref{sec:results}).

\section{The Karlsruhe Tritium Neutrino Experiment} \label{sec:KATRIN}
In the following paragraphs, the relevant components of the KATRIN experiment and the processes within the source are described.

\subsection{Experimental Setup}

The KATRIN experiment is located at the Karlsruhe Institute of Technology, which operates the Tritium Laboratory Karlsruhe (TLK). This facility is capable of providing the large amounts of high-purity tritium required to operate the tritium source of the KATRIN experiment \cite{Sturm2021}. 

Molecular tritium is supplied to the windowless gaseous tritium source (WGTS) \cite{Heizmann_2017} from a dedicated tritium circulation loop system \cite{Hillesheimer2024} at a rate of about \SI{40}{\gram\per\day} and at a purity of $>\SI{95}{\percent}$ \cite{tritiumpurity}. Half of the electrons of the isotropic $\upbeta$-decay are guided in a cyclotron motion by a solenoid field at about $B_{\mathrm{src}}=\SI{2.5}{\tesla}$ in the direction of the main spectrometer (downstream direction). The tritium gas is pumped out by the differential and cryogenic pumps of the transport section, achieving a total reduction of the tritium flow along the beamline by 14 orders of magnitude \cite{Marsteller2021, Roettele2023}. The tritium source tube is operated at $T_{\mathrm{src}}=\SI{80}{\kelvin}$, as this is a temperature that allows for the co-circulation of meta-stable $\kr$ \cite{Marsteller2022}.

To filter the electron energy with sufficient precision, the MAC-E filter principle \cite{Beamson1980} is applied in the main spectrometer.
The strong magnetic field in the source drops to $B_\text{ana}$ (which is on the order of $\SI{100}{\micro\tesla}$) in the so-called analyzing plane in the main spectrometer. The conservation of the electrons' orbital magnetic moment 
 leads to a collimation of the electrons' momenta into the forward direction. At the same time, the electrons run up an electrostatic retarding voltage $U$, which can only be passed by electrons with sufficient energy in forward direction. Electrons can still carry some energy, associated with the transverse momentum relative to the magnetic field, coming from not-collimated momentum, which leads to a transmission width of the MAC-E filter of 
\begin{equation}
   \Delta E= E\frac{B_\text{ana}}{B_\text{max}}\frac{\gamma+1}{2},
   \label{eq:TransWidth}
\end{equation}
where $\gamma$ is the electron Lorentz factor \cite{Kleesiek2019}. The transmission of electrons through the spectrometer depends on the energy and pitch angle in the source, the reduction of magnetic field towards the analyzing plane, and the retarding voltage. Studies have shown that there is a trade-off between a low magnetic field in the analyzing plane, which enables high resolution, and higher fields, which reduce background electron rates \cite{Trost2019_1000090450}. Additionally, electrons with high starting angles are reflected by the magnetic mirror effect, which is determined by the pinch magnetic field $B_\text{max}$ in front of the detector. 
Electrons fulfilling the transmission conditions are counted at the focal plane detector (FPD) \cite{Amsbaugh2015}. 
The segmentation of the FPD into 148 pixels provides a spatial resolution of the counted electrons.
The integral $\upbeta$-spectrum is then scanned by a stepwise variation of the retarding potential energy $qU$, with $q=-e$ as the electron charge.

In the isotropic tritium source, half of the electrons are guided towards the rear wall (RW) of the source (upstream direction).
The rear wall is a gold-coated stainless steel disc. A bias voltage is applied to it, thereby influencing the source potential.

\begin{figure*}
    \centering
    \includegraphics[width=\textwidth]{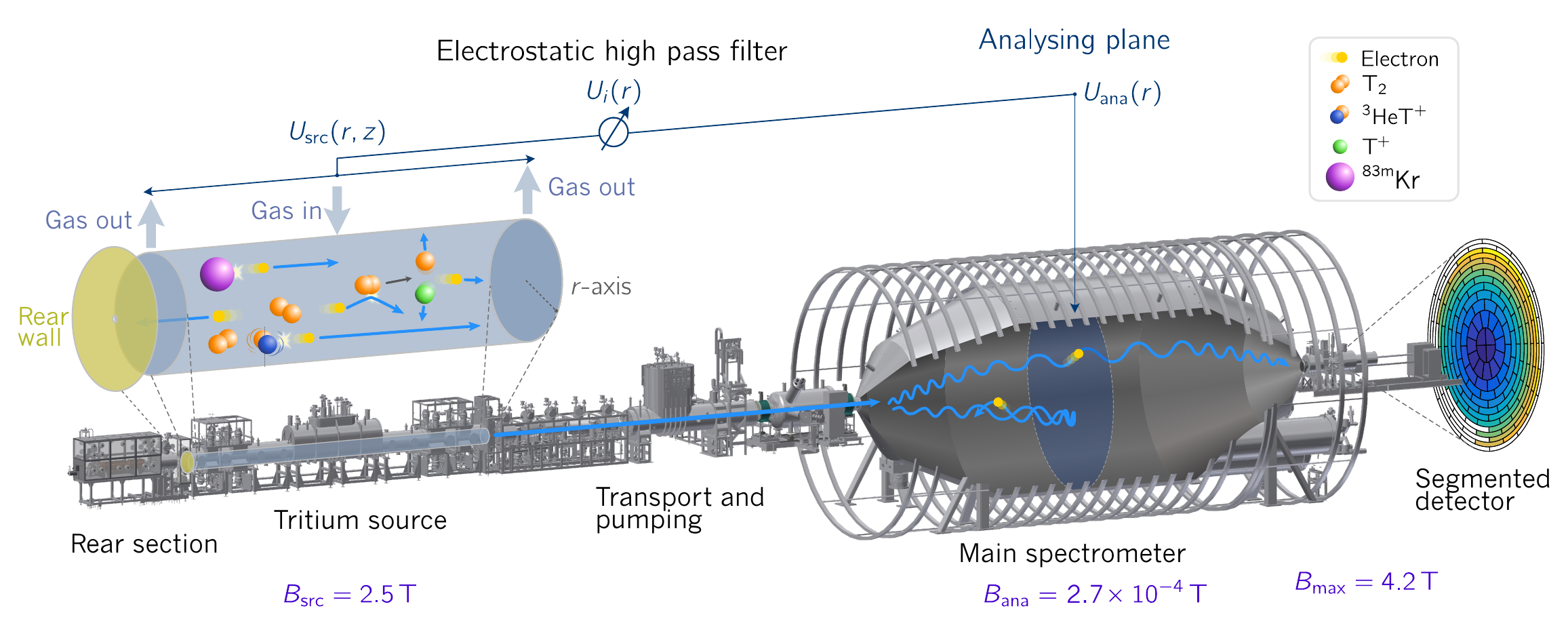}
    \caption{The KATRIN beamline. Tritium is circulated through the source section. The $\upbeta$-decay electrons from the tritium decay are guided through the transport and pumping section into the main spectrometer, where an integral measurement of their energy is performed. The spectrometer follows the MAC-E filter principle, where the electrons' momenta are magnetically collimated parallel to an electrostatic retarding field. Electrons with surplus energy can pass and are counted at a detector. The retarding potential is defined by the difference of the spectrometer ($U_\text{ana})$ and source ($U_\text{src}$) potentials. The latter is influenced by the surface of the walls, a bias voltage on the rear wall, and by a cold plasma formed by scattering of the $\upbeta$-decay electrons on the gas molecules. Gaseous $\kr$ can be co-circulated with tritium to study the source potential.}
    \label{fig:beamline}
\end{figure*}

\subsection{Formation of a cold plasma and electric potential of the tritium source}

In the KATRIN source, tritium gas is injected at constant flow into the beam tube (BT) center. At both ends of the beam tube, the gas is pumped off, thereby forming a stable tritium column density (CD), that is the number of molecules per area when integrating over the source length.

A large fraction of $\upbeta$-electrons undergoes scattering on the source gas, leading to energy losses, dissociation, and ionization. Every primary decay electron generates about 10 times more secondary particles \cite{Nastoyashchii2005}. The resulting number density of charged particles of \num{e11} to \SI{e12}{\per\cubic\meter} makes the KATRIN source a low-density, low-temperature plasma \cite{Kuckert2016}. A detailed description needs to consider the surface potentials, the properties of the neutral gas, geometries, and the confinement of the charges by the magnetic field.

Substantial efforts have been made to simulate these underlying processes in high detail. The plasma to be described is classified as bound, strongly magnetized, partly collisional, and partly ionized \cite{Kellerer2022SimulationMethods}. For the KATRIN goal, these simulations need to provide the spatial potential distribution and insight into plasma instabilities. 
Initial studies assumed a diffusive plasma with total thermalization
of all charged particles in the source. 
A. F. Nastoyashchii et al.~\cite{Nastoyashchii2005} approached the question of the WGTS plasma with Monte Carlo simulations, in which only the mean electron spectrum for the entire source was derived. L. Kuckert used a drift-diffusion fluid-dynamical approach \cite{Kuckert2016}.

Recently, J. Kellerer and F. Spanier employed a two-stage simulation methodology, designed to more accurately model the properties of plasma within the source. 
Atomic interactions, specifically collisions of charged particles with neutral gas, are modeled within a Monte Carlo framework \cite{Kellerer2022}. Its output is subsequently integrated into particle-in-cell simulations to account for plasma effects \cite{Kellerer2022SimulationMethods}. These computationally expensive simulations reveal the presence of a non-thermal transitional energy range, distinct from the $\upbeta$-electrons and thermalized electrons within the electron energy spectrum, which exerts a significant influence on the properties of the plasma.

The simulations predict a longitudinal inhomogeneity of the potential on a \SI{10}{mV} scale. However, the lack of knowledge regarding the input parameters for the simulation, such as the boundary conditions at the rear wall and the source's permittivity, hinders the precise prediction of this value. To circumvent this source of uncertainty, we apply a phenomenological approach of direct measurement of the source-poten\-tial inhomogeneity's effect on the $\upbeta$-spectrum (subsection \ref{sec:SourceObservables}), and we minimize that inhomogeneity by adjusting the rear-wall voltage (subsection \ref{sec:Opt_RW_definition}).

\subsection{Observables of the source potential and their impact on the neutrino mass}\label{sec:SourceObservables}

\begin{figure*}
    \centering
    \includegraphics[width=.85\textwidth]{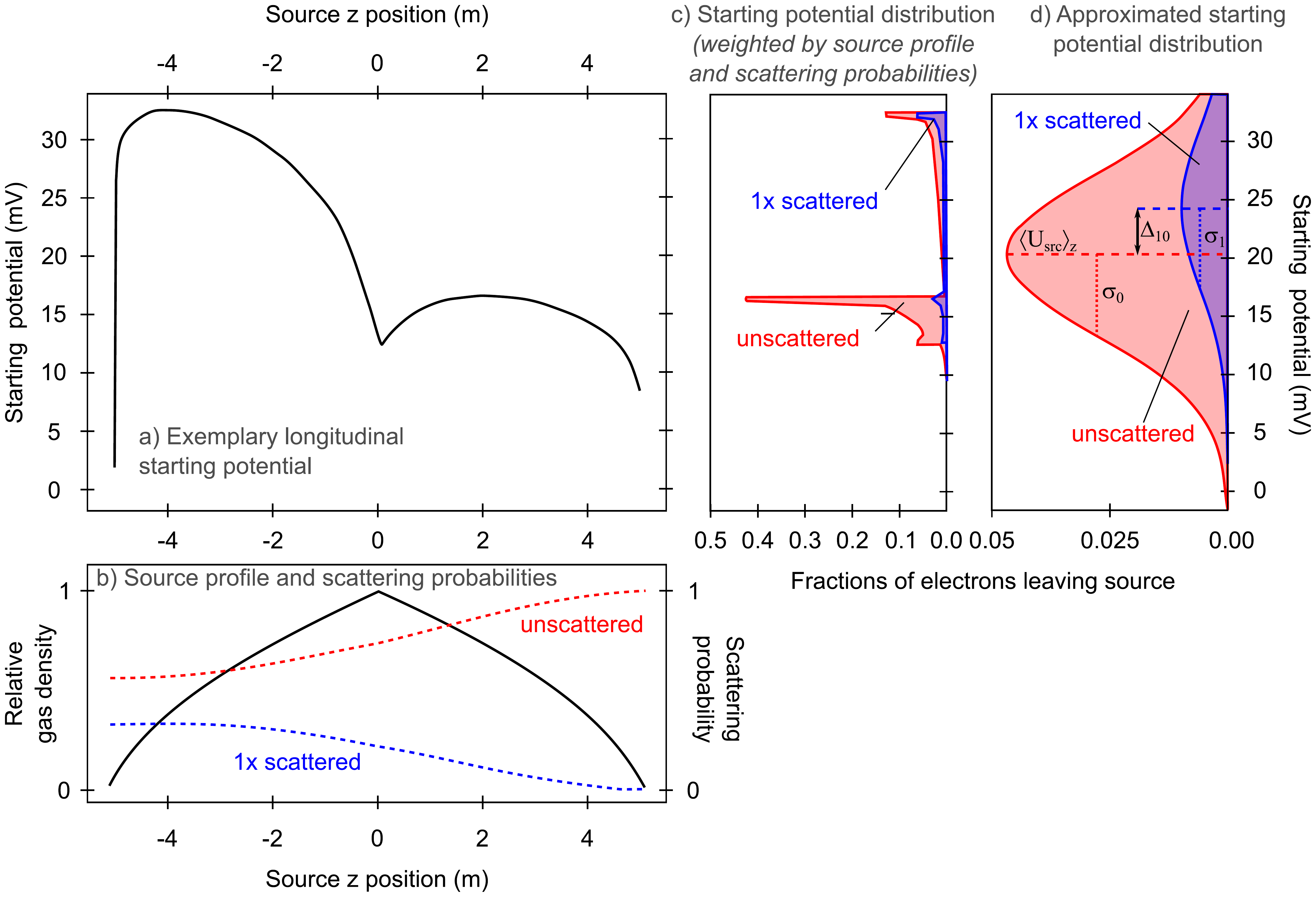}
      \caption{Visualization of the relation between longitudinal starting potential and measurable observables. $z=0$ corresponds to the gas injection point of the tritium source. The positive $z$-axis points towards the KATRIN spectrometer (upstream), while the negative $z$-axis points towards the rear wall. (a)~Exemplary longitudinal starting potential $U_\mathrm{src}(z,r,\phi)$ for arbitrary values of $r$ and $\phi$, according to \cite{Kuckert2016}. b)~Longitudinal gas profile (solid line) and scattering probability for electrons leaving the KATRIN source in downstream direction without scattering (dashed red line) or with a single scattering event (dashed blue line). The plot shows that unscattered electrons preferentially originate from the downstream part of the source while singly scattered electrons originate largely from the upstream part. Higher order scattering is not displayed. c)~Electron starting potential distribution in the KATRIN source, considering the source gas profile and the scattering probabilities, for both unscattered and singly scattered electrons \cite{Machatschek2021}. d)~For illustration purposes: Starting potential distribution approximated by a Gaussian for unscattered and single scattered electrons. The average potential is described by $\langle U_\text{src}\rangle_z(r,\phi)$, the variance of distributions by $\sigma_0^2$ and $\sigma_1^2$, indicating the source-potential broadening. The shift between the distributions is denoted by  $\Delta_{10}$ and accounts for asymmetries in the potential \cite{Machatschek2021}.  }
    \label{fig:sketch-on-starting-potential}
\end{figure*}

While the segmented detector resolves the measured spectrum in the $(r,\phi)$ plane in cylinder coordinates, a variation of the source potential in the $z$ direction is averaged over the \SI{10}{m} long WGTS beam tube. The $z$-averaged source potential $\langle U_\text{src}\rangle_z(r,\phi)$ is obtained from integration over $U_\text{src}(r,\phi,z)$ weighted with the $z$-dependent gas distribution and scattering probability, shown in figure~\ref{fig:sketch-on-starting-potential}.

However, a longitudinal inhomogeneity of the source potential leads to a distribution of the starting electron energy, which has to be convolved with the energy spectrum. At leading order, a longitudinal inhomogeneity is described by a Gaussian broadening parameter $\sigma_0$, which is the standard deviation of the source potential energy of the unscattered electrons. This observable has been introduced in a similar way in the investigations of \cite{Belesev2007InvestigationExperiment}. A detailed illustration of the phenomenological model of the source-potential observables is shown in figure \ref{fig:sketch-on-starting-potential}.

Scattered electrons predominantly originate from the rear part of the source and unscattered electrons from the front part. Hence, a second observable $\Delta_{10}$ describes the difference in mean potential energy of singly-scattered and unscattered electrons. It is a measure of the asymmetry of the potential with respect to the central injection point. The numerical indices $0$ and $1$ indicate the number of scatterings an electron has undergone. While the zeroth order is strongly dominant, this description can be extended to electrons of multiple scatterings, and an effective value $\Delta_\text{P}$ combining all relevant scatterings can be constructed\footnote{This is not necessary for the source-potential broadening: The $\sigma_\mathrm{i>0}$ can be simply approximated by $\sigma_0$, since the source-potential inhomogeneity is negligible compared to the width of the energy-loss distribution of scattered electrons.}. More details are provided in \cite{Machatschek2021}.

If not taken into account, the inhomogeneous source potential leads to a bias of the squared neutrino-mass result \cite{Machatschek2021}
\begin{align}
    \Delta m_\upbeta^2 &=-2(\sigma_0^2+\sigma_\text{add}^2)-\sum_{i>0}\epsilon_i \Delta_{i0}\\
    &=-2\sigma^2-\epsilon \Delta_\text{P}.\label{eq:plasma-parameters}
\end{align}
Here the parameter $\sigma_\text{add}^2$ contains additional broadening contributions, such as hypothetical temporal fluctuations. $\epsilon$ describes the susceptibility of the squared neutrino mass $m_\upbeta^2$ to the asymmetry parameter $\Delta_\text{P}$, and is on the order of \SI{1}{eV}. It depends on the fraction of scattered electrons and is determined from Monte Carlo simulations \cite{Machatschek2021}. Given that the expected inhomogeneity of the potential is on a \SI{10}{mV} scale, the second term in equation~\ref{eq:plasma-parameters} is strongly dominant. If the source-potential observables vary radially or with time, equation~\ref{eq:plasma-parameters} holds for the respective averages.

Since $\Delta_{i0}$ is a measure of the inhomogeneities caused by asymmetric potentials and $\sigma_0$ a measure of the overall longitudinal inhomogeneity, $\sigma_0$ allows setting an upper limit \cite{Machatschek2021}:
\begin{equation}
\label{equ:delta10}
    |\Delta_{i0}| \leq \kappa_i \sigma_0.
\end{equation}
The parameter $\kappa_i$ is related to the overlap of the spatial distributions of scattered and unscattered electrons. $\kappa_1$ is estimated to be 0.69 for \SI{75}{\percent} column density of tritium and 0.62 for \SI{40}{\percent} column density, with respect to a nominal column density of \num{5e17} molecules per cm$^2$. Equation~\ref{equ:delta10} holds for the longitudinal component $\sigma_0$ of the total broadening $\sigma$; However, since $\sigma$ is always larger than $\sigma_0$, it is conservative to use a measured value $\sigma$ to constrain $\Delta_{i0}$ by this equation. An analogous equation with a combined coefficient $\kappa$ holds for $\Delta_\text{P}$. The $\kappa$ values are reported in \cite{Machatschek2021}.

By determining $\sigma$ and $\Delta_\text{P}$ and considering them as model inputs of the neutrino-mass analysis, the neutrino-mass bias as predicted by equation \ref{eq:plasma-parameters} can be circumvented. The design goals for KATRIN stipulate uncertainty contributions on $m_\upbeta^2$ by individual systematic effects below $\SI{7.5e-3}{\electronvolt\squared}$ \cite{BEGLARIAN2005}. To meet this requirement $\Delta_{10}$ needs to be determined with a precision of a few meV. This is done with experimental campaigns using $\kr$, which allow for an in-situ measurement of the parameters $\sigma_0$ and $\Delta_{10}$.

\subsection{Influence of the rear-wall voltage on the source potential}
\label{sec:Opt_RW_definition}

For neutrino-mass measurements, it is preferable to operate the source in a setting with minimal radial inhomogeneity of the source potential, in order to simplify the neutrino-mass analysis. Therefore, a simple empirical approach for the radial dependence of the source potential is made. 

Energies measured with the KATRIN setup are shifted from the true value by the difference of the main spectrometer (MS) work function $\Phi_\text{MS}$ and the source potential energy. 
Neglecting an azimuthal dependence of the source potential, the observed energy of unscattered electrons is given by
\begin{equation}
E_\text{obs,0}(r)=E_\text{true}-\Phi_\text{MS}+q\langle U_\text{src}\rangle_z(r).
\end{equation}
Plasma simulations, for instance those conducted in \cite{Kuckert2016}, predict that
the source volume adopts mainly the electric potential energy of the rear wall $\Phi_\text{RW}+qU_\text{RW}$, where $\Phi_\text{RW}$ is the mean work function of the rear wall and $U_\text{RW}$ is a bias voltage applied to the rear wall. This coupling of the plasma potential to the rear wall is only fully realized if the bias voltage compensates the difference of $\Phi_\text{RW}$ to the mean work function $\Phi_\text{BT}$ of the grounded cylindrical beam-tube walls: $-qU_\text{RW}\stackrel{!}{=}\Phi_\text{RW}-\Phi_\text{BT}=:\Delta\Phi_\text{RW,BT}$. Otherwise, outer parts of the source volume would exhibit weaker coupling to the rear wall, given their proximity to the beam-tube walls. These considerations, combined with observations outlined in \cite{Friedel2020}, motivate the following simplified empirical model of $\langle U_\text{src}\rangle_z(r)$: 
\begin{align}
	q\langle U_\text{src}\rangle_z(r)&=c(r)(\Phi_\text{RW}+qU_\text{RW})+(1-c(r))\Phi_\text{BT}\\
    \label{eq:source-pot-model}
	&=c(r)(qU_\text{RW}+\Delta\Phi_\text{RW,BT})+\Phi_\text{BT}.
\end{align}
The so-called coupling coefficient $0\leq c(r)\leq1$ depends on the radial distance to the beam axis $r$.
 By adjusting the rear-wall voltage such that $-qU_\text{RW}$ is equal to the difference in work function between the rear wall and the beam tube, the coupling coefficient in equation~\ref{eq:source-pot-model} vanishes. This setting is called the optimal rear-wall voltage $U_\text{RW,opt}$. At this specific rear-wall voltage, the radial inhomogeneity of the potential is minimized. However, it is anticipated that a correlation exists between the radial and longitudinal ($z$) inhomogeneities. As a result, this configuration suggests a near-minimal longitudinal inhomogeneity.
Drifts of the rear-wall and beam-tube work functions lead to a change of the optimal rear-wall voltage over the course of weeks, which is monitored by calibration measurements and accounted for in the data analysis.

\section{Krypton measurements for source-potential investigations}
\label{sec:Kr_meas_campaign}

$\kr$ is widely used as a calibration source in particle physics experiments due to its quasi-monoenergetic conversion-elec\-tron lines and its fast decay, meaning that it does not contaminate the experiment. At KATRIN, its gaseous form at standard measurement conditions is vital to determine the spatially extended potential of the source.

The following paragraphs describe the conversion-elec\-tron spectrum of the $\kr$ isotope and the application of gaseous $\kr$ in the calibration measurement campaigns.

\subsection{$\kr$ conversion electrons and suitable lines}

$\kr$ is produced by the electron capture of $^{83}$Rb with a half-life of \SI{86.2}{d}. The metastable isotope decays with a half-life of \SI{1.83}{h} in a \SI{32.2}{keV} nuclear $\gamma$ transition to an intermediate state of \SI{9.4}{keV}, and then via a second $\gamma$ transition with a half-life of \SI{155}{ns} into the ground state \cite{Venos2018}.
Both transitions are highly converted and exhibit a spectrum of quasi-monoenergetic conversion-electron lines. The mean energy $\mu$ of the conversion electron is given by
\begin{equation}
    \mu=E_\text{trans}-E_\text{rec}-E_\text{bind},
\end{equation}
where $E_\text{trans}$ is the nuclear transition energy, $E_\text{rec}$ is the recoil of the conversion-electron emission, and $E_\text{bind}$ is the binding energy of the subshell the electron was emitted from.

For calibration purposes at KATRIN, conversion-elec\-tron lines from the \SI{32.2}{keV} transition are used, as they exhibit energies above the endpoint of the tritium $\upbeta$-decay spectrum.
Two lines are the most suitable: First, the L$_3$-32 line offers high intensity. Measurements of its line position allow for fast investigations of source-potential shifts. However, its $\approx \SI{1}{eV}$ natural line width \cite{Venos2018}, given by the lifetime of the vacancy in the electron shell, limits its applicability to investigate the energy broadening. To assess the source-potential-related broadening of a few meV magnitude through fitting this relatively broad peak, it is essential to ascertain the relative accuracy of the Lorentzian line width at the $10^{-3 }$ level, which is highly challenging \cite{Machatschek2021}.

\begin{figure}
     \centering
     \includegraphics[width=.5\textwidth]{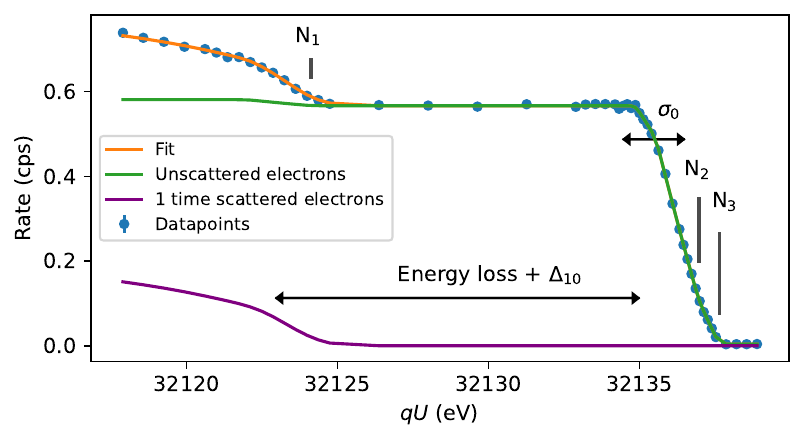}
     \caption{Integral spectrum of the N$_{23}$-32 line doublet and the N$_1$-32 line of $\kr$ from internal conversion. The spectrum consists of an unscattered portion (green) and a singly scattered portion (purple) which is shifted by approximately \SI{13}{eV} towards lower energies due to inelastic energy losses. The inhomogeneous source potential is characterized by a line broadening $\sigma_0$ and an additional shift $\Delta_{10}$ between the spectra of singly scattered and unscattered electrons. Both parameters can be directly determined from the spectrum of $\kr$. The region of singly scattered electrons from N$_{23}$-32 overlaps with the unscattered N$_1$-32 line.
     }
     \label{fig:KrSpec}
 \end{figure}

The lines of the N$_{23}$-32 doublet have negligible natural width, allowing for precise determination of broadenings. This is because the N$_3$ subshell is the outermost occupied subshell in the $\kr$ atom, meaning that the vacancy of the conversion-electron emission could only be refilled by electrons from the surrounding gas in the WGTS. Estimates of the interaction rate with atoms or free electrons in the source yield an expected line width on the order of \SI{e-10}{eV}, which is negligible.
The natural line width of the N$_2$-32 line is equally negligible, since the filling of the N$_2$ shell by N$_3$ electrons is suppressed by the corresponding selection rules. As the $\ntwothree$ doublet has an intensity approximately 50 times lower than L$_3$-32 \cite{Venos2018}, a high source strength is needed for measuring these lines. The N$_{23}$-32 doublet is accompanied by the N$_1$-32 line and nearby shake lines. The integral line spectrum for the N$_{123}$-32 lines as measured at KATRIN is shown in figure \ref{fig:KrSpec}. 

The analysis of the N$_{123}$-32 spectra should directly yield the source-potential broadening $\sigma_0$ and the asymmetry parameter $\Delta_{10}$, but needs to consider systematic effects. Currently, the N$_1$-32 and shake-line parameters are determined in dedicated measurements at KATRIN. They are required for the precise determination of the asymmetry parameter $\Delta_{10}$ because these lines overlap with the region of singly scattered electrons from N$_{23}$-32. Furthermore, the energy loss of \SI{32}{keV} electrons inelastically scattering on T$_2$ is currently under investigation. In a reassessment of the recorded data we aim to directly determine the asymmetry parameter $\Delta_{10}$ in a future publication. This work focuses solely on determining the broadening of the N$_{23}$-32 lines, and $\Delta_{10}$ is constrained using equation \ref{equ:delta10}.

\subsection{The $\kr$ calibration campaigns KrM5 and KrM9}

Neutrino-mass measurement campaigns have been recorded using different tritium column densities, which, in principle, can change the source plasma and thus the electric potential. Consequently, $\kr$ campaigns have been carried out under various source conditions, shown in table \ref{tab:measurements}.
In this work, two dedicated measurement campaigns using $\kr$ are discussed.\\

First, the KrM5 campaign was conducted right after the fifth neutrino-mass measurement campaign KNM5 \cite{KMN1-5-paper}. $\kr$ was produced using a high-activity $^{83}$Rb source with an initial activity of $\approx\SI{10}{GBq}$, produced by NPI Řež.
At the beginning of KrM5, a measurement of the L$_3$-32 line was carried out at varying rear-wall voltages over the course of 5.5 days (table~\ref{tab:measurements}, measurement a). The column density of co-circulated tritium was \SI{75}{\percent} of the design value of $\num{5e17}$ molecules per cm$^2$, and is the standard for neutrino-mass measurements since KNM3-NAP (cf. table~\ref{tab:FinalInputs}) \cite{KMN1-5-paper}. This measurement allowed the determination of the optimal rear-wall voltage set-point, as defined in subsection \ref{sec:Opt_RW_definition} and obtained in subsection~\ref{sec:L3_75percent}. This voltage was then applied to the following measurements:

A dedicated high-statistics measurement of the N-32 lines was performed at \SI{75}{\percent} column density to precisely determine the broadening of the source potential $\sigma$ (table~\ref{tab:measurements}, measurement b).
Over the course of 25 days, 180 spectra containing the energy-loss region for the singly scattered electrons were recorded.

The measurement of the N-32 lines was repeated at a column density of \SI{40}{\percent} to determine the dependence of the source-potential broadening on the column-density setting (table~\ref{tab:measurements}, measurement c). The data set consists of 16 spectra recorded over the course of one day. The shorter measurement time in the \SI{40}{\percent} setting is sufficient, since the krypton activity in the source is increased by a factor of around 500 in this setting with respect to the \SI{75}{\percent} setting \cite{Marsteller2022}.

While neutrino-mass measurements are preferably performed at the optimal rear-wall voltage, drifts of the surface potential of the rear wall and the source beam-tube walls may occur.
In order to investigate the rear-wall voltage dependence of the source-potential observables, the N-32 lines were measured at different rear-wall voltages, once at \SI{40}{\percent} column density and once at \SI{75}{\percent} column density (table~\ref{tab:measurements}, measurements d and e).

Two years after KrM5, another set of krypton spectra was taken in the KrM9 campaign, following the KNM9 neu\-tri\-no-mass measurements (table~\ref{tab:measurements}, measurements f and g). The $^{83}$Rb source activity was around \SI{9}{GBq}.
The work functions of the surfaces within the source, the beam tube and the rear wall, underwent drifts in the intermediate time span. Moreover, the rear-wall surface was subjected to two intensive ultraviolet/ozone cleaning sessions in the interval between KrM5 (June to August 2021) and KrM9 (May 2023) \cite{AkerRearWall24}. Consequently, an effective shift of the optimal rear-wall voltage of \SI{0.8}{\volt} was observed, which required a corresponding adjustment in the set voltage of the rear wall. The N-line region was scanned at a column density of \SI{75}{\percent} for another 15 days, accumulating 97 scans. Furthermore, a one-day measurement was performed with 10 scans of the N-line region at a column density of \SI{40}{\percent}.

The analysis of all data sets is described in detail in section \ref{sec:Ana}.

\begin{table}[]
 \caption{Overview of settings for $^{83\text{m}}$Kr conversion-electron line measurements for characterization of the source-potential systematics at KATRIN. The top part of the table contains measurements from the KrM5 campaign, and the bottom part measurements from the KrM9 campaign. The various measurements differ in the scanned line, tritium column density and rear-wall voltage.}
    \label{tab:measurements}
    \centering
    \begin{tabular*}{\columnwidth}{@{\extracolsep{\fill}}lcccc@{}}
    \toprule
       Measurement  & Line & CD & RW Voltage & Duration \\\midrule
       a & L$_3$-32 & \SI{75}{\percent} & -0.6\,V to 0.2\,V & \SI{5.5}{d} \\
       b & N$_{23}$-32 & \SI{75}{\percent} & -0.3\,V  & \SI{25}{d} \\
       c & N$_{23}$-32 & \SI{40}{\percent} & -0.3\,V  & 1\,d \\
       d & N$_{23}$-32 & \SI{40}{\percent} & -0.6\,V to 0.2\,V  & 1.5\,d \\
       e  & N$_{23}$-32 & \SI{75}{\percent} & -0.6\,V to 0.2\,V & 1.5\,d \\\midrule
       f & N$_{23}$-32 & \SI{75}{\percent} & 0.5\,V  & 15\,d \\
        g & N$_{23}$-32 & \SI{40}{\percent} & 0.5\,V  & 1\,d \\\bottomrule
    \end{tabular*}
   
\end{table}

\section{Data analysis}\label{sec:Ana}

In order to infer the optimal rear-wall voltage $U_\text{RW,opt}$, the source-potential broadening $\sigma_0$, and the asymmetry parameter $\Delta_{10}$ from the recorded $\kr$ line spectra, a model of the line shapes and the transmission properties of the experiment is fit to the data.

To account for the broadening, the natural Lorentzian line shape with line position $\mu$ and width $\Gamma$ is convolved with a Gaussian shape of variance $\sigma^2$, giving the Voigt profile $V(E,\mu,\sigma^2,\Gamma)$. Scaled by the line amplitude $A$ one obtains the differential rate of a single line $D(E, A, \mu,\sigma^2,\Gamma)$.
The differential spectrum of multiple lines consists of the sum of Voigt profiles.

The L$_3$-32 line has a line position of $\mu=\SI{30472.2\pm0.5}{\electronvolt}$ and an intrinsic width of $\Gamma=\SI{1.19\pm0.24}{\electronvolt}$ \cite{Venos2018}. The N$_{23}$-32 line doublet consists of the N$_2$-32 line at \SI{32136.7\pm0.5}{\electronvolt} and the N$_3$-32 line at $+\,\SI{0.670\pm0.014}{\electronvolt}$ relative to the N$_2$-32 line \cite{Venos2018}. For the data analysis, the natural line widths of N$_2$-32 and N$_3$-32 are assumed zero, as described in sec.~\ref{sec:Kr_meas_campaign}.
The modeling of the integrated spectrum measured with the KATRIN main spectrometer is based on \cite{Kleesiek2019}.
The integrated spectrum $\mathcal{R}$ of the krypton conversion-electron lines can be written as

\begin{equation}
\label{eq:krypton_full_model}
\mathcal{R}(qU)=\int\limits_{0}^{\infty}  \! \mathrm{d}E  \int\limits_{0}^{\theta_\mathrm{max}}  \! \mathrm{d}\theta ~ \mathcal{T}(E,qU,\theta)\sin{\theta} P_0(\theta, \rho d) D(E) + \mathcal{R}_\mathrm{bg}.
\end{equation}
The differential spectrum $D$ is convolved with the experimental response, reflecting the transmission properties $\mathcal{T}$ of the MAC-E filter, as well as inelastic scattering of electrons on tritium molecules in the source by $P_0(\theta, \rho d)$. The functions are convolved with respect to the pitch angle $\theta$ of the electrons relative to the magnetic field line, coming from the isotropic angular distribution of the source. The maximum pitch angle $\theta_\text{max}$ with which an electron can pass the main spectrometer is defined by the magnetic reflection at the pinch magnet, $\theta_\text{max}=\arcsin\sqrt{\frac{B_\text{src}}{B_\text{max}}}$. The transmission condition for the electrons is
\begin{equation}
    \mathcal{T}(E,qU,\theta)=\begin{cases}
1 & \, \text{if} \,\, E\left(1-\sin^2\theta \frac{B_\text{ana}}{B_\text{src}}\frac{\gamma+1}{2}\right) > qU, \\
0 & \, \text{else}.
\end{cases}
\end{equation}
Electrons have a probability $1-P_0(\theta,\rho d)$ for inelastic scattering in the source at a column density $\rho d$. Although the analysis interval mainly selects unscattered electrons\footnote{The analysis window contains a tiny fraction of electrons of other krypton lines that have scattered inelastically off tritium. They were shown to have a negligible impact on the determination of the line broadening and position.}, the pitch-angle dependence of the zero-scattering probability modifies the shape of the integrated spectrum. The scattering probabilities are calculated by assuming the scattering to be a Poisson process\footnote{Deviations from the Poisson distribution have been investigated and found to be negligible.}, and by incorporating simulations of the tritium and krypton gas distributions in the source \cite{Machatschek2021}.
An additional correction of the spectrum model accounts for the angular-dependent detection efficiency of the FPD, which is modeled by a second-order polynomial expansion in $\theta$ \cite{KMN1-5-paper}. Furthermore, energy losses of electrons due to synchrotron radiation in the source and transport section occur and are considered in the analysis \cite{Kleesiek2019}. 
Finally, a $qU$-independent background $\mathcal{R}_\text{bg}$ is added to the model.

Simulations of the electromagnetic fields in the spectrometer are incorporated in the analysis, in order to account for the inhomogeneities of the electric potential in the analyzing plane and of the maximum magnetic field. The analyzing plane magnetic field is treated as a nuisance parameter, except where noted differently.

In accordance with the tritium $\upbeta$-analysis, 22 pixels shadowed by structural components of the beamline or with high intrinsic background \cite{KMN1-5-paper} are excluded from the analysis of the $\kr$ data.

\subsection{L$_3$-32 measurements for the determination of the optimal rear-wall voltage}
\label{sec:L3_75percent}

\begin{figure}
    \centering
    \includegraphics[width=.5\textwidth]{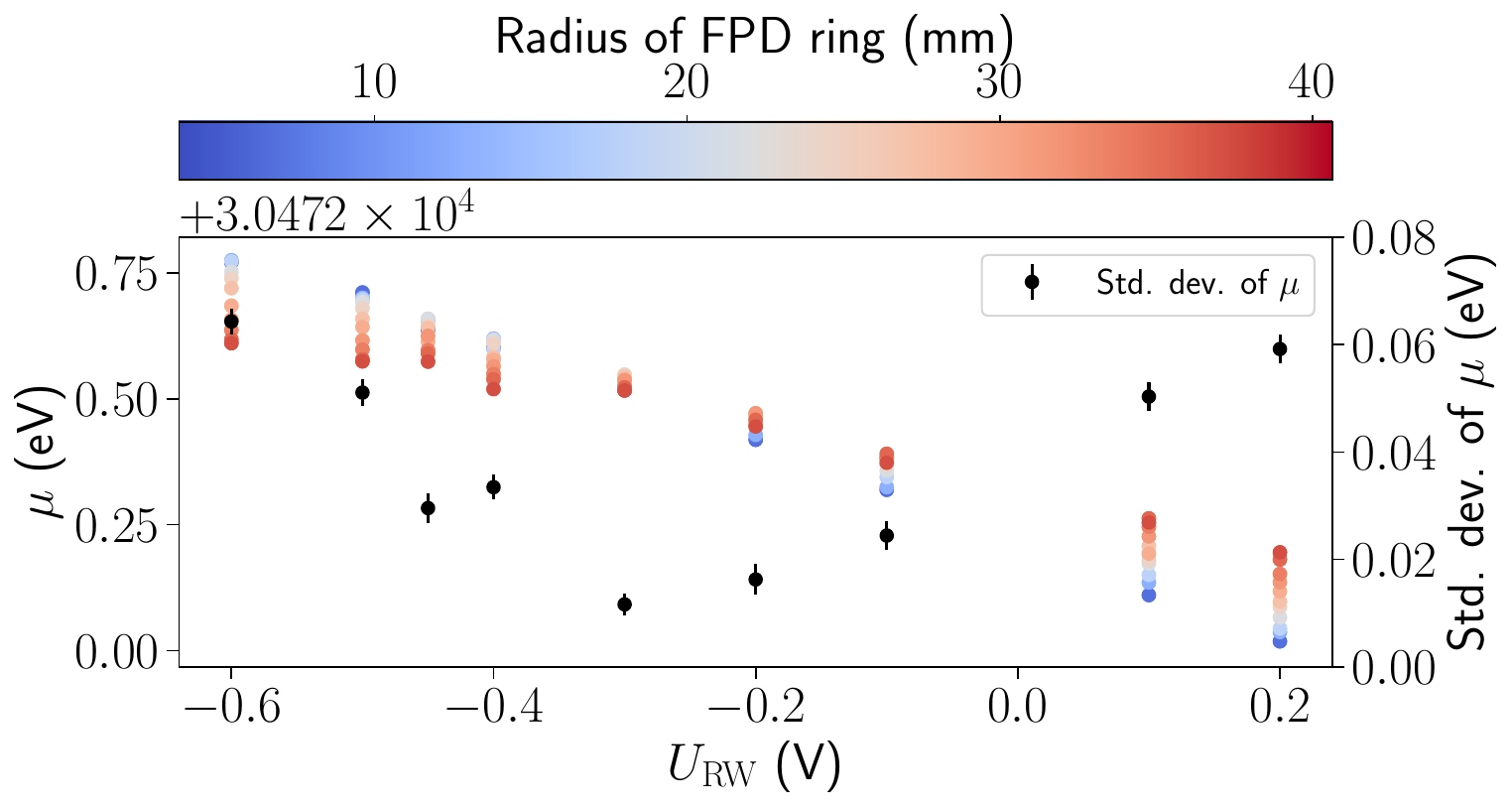}
    \caption{Measurement of the L$_3$-32 line position $\mu$ as function of rear-wall voltage $U_\mathrm{RW}$. The line position (colored) depends on both the rear-wall voltage and the radial distance from the source tube center, as measured by the radius of the rings at the focal plane detector. The standard deviation of ring-wise line positions (black) per rear-wall voltage set point is minimal around \SI{-0.3}{V}, which was identified as the optimal rear-wall voltage and used for the subsequent measurements.}
    \label{fig:L3muVsURW}
\end{figure}

In the KrM5 campaign, measurements of the L$_3$-32 line were performed at nine different rear-wall voltages to obtain the optimal rear-wall voltage. After every 12 hours $U_\text{RW}$ was changed. In the analysis, the FPD pixels are grouped into 11 concentric rings of 4-12 pixels each to more easily study the radial dependence of the source potential. The $B_\text{ana}$ values are set to the average simulated values in the respective rings. For simplicity the Lorentzian width is treated as a nuisance parameter to include all broadenings, while the Gaussian broadening component of the spectrum is neglected. Figure~\ref{fig:L3muVsURW} shows the distribution of ring-wise line positions as function of $U_\text{RW}$. The radial spread of line positions is minimal at $U_\text{RW}\approx\SI{-0.3}{V}$, which was adopted as the optimal rear-wall voltage for the subsequent KrM5 measurements.

\subsection{N-32 measurements for the determination of the source-potential broadening}
\label{sec:NAnalysis}

The differential spectrum of the N$_{23}$-32 doublet is given by a sum of Voigt functions,
\begin{equation}
\begin{split}
D(E)=&A_{\text{N}_2}\cdot\Bigl[ V\left(E,\mu_{\text{N}_2},\sigma^2,\Gamma=0 \right)\\&+ R_{\text{N}_3,\text{N}_2}\cdot V\left(E,\mu_{\text{N}_2}+\Delta \mu_{\text{N}_3,\text{N}_2}, \sigma^2,\Gamma=0\right)\Bigr],
\end{split}
\end{equation}
where $A_{\text{N}_2}$ and $\mu_{\text{N}_2}$ are the amplitude and line position of N$_2$-32, $R_{\text{N}_3,\text{N}_2}$ is the amplitude ratio of N$_3$-32 to N$_2$-32 and $\Delta \mu_{\text{N}_3,\text{N}_2}$ is their distance.
The squared Gaussian line broadening $\sigma^2$ is common for all lines in the spectrum.
To obtain this broadening, the fit interval was set to a \SI{6}{eV} range around the N$_{23}$-32 doublet.

\begin{table}[h]
    \centering
    \caption{Additional broadenings due to fluctuations of the high voltage and the Doppler effect, applicable for the N-32 line measurements at the optimal rear-wall voltage. The uncertainties of all other broadening contributions lie below \SI{e-6}{eV^2} and are neglected.}
    \begin{tabular*}{\columnwidth}{@{\extracolsep{\fill}}lcc@{}}
    \toprule
        &  \SI{40}{\percent} CD &  \SI{75}{\percent} CD  \\
        \midrule
       $\sigma_\text{HV}^2$ (\SI{e-3}{eV^2}) in KrM5 & \num{0.3+-0.1} & \num{0.21+-0.05}\\
       $\sigma_\text{HV}^2$ (\SI{e-3}{eV^2}) in KrM9 & \num{0.18+-0.07} & \num{0.16+-0.04}\\\midrule
       $\sigma_\text{Doppler}^2$ (\SI{e-3}{eV^2}) & \multicolumn{2}{c}{\num{2.98+-0.06}} \\
        \bottomrule
    \end{tabular*}
    \label{tab:HV_broadenings}
\end{table}

Simulations on Asimov data \cite{AsimovPaper} were performed determining the impact of the systematic uncertainties, including the influences of the magnetic fields and the parameters of the neighboring $\kr$ shake and N$_1$ lines, thereby assuring their proper treatment in the analysis. The largest systematic effects stem from uncertainties in $B_\text{ana}$ and $B_\text{src}$. Thus, $B_\text{ana}$ is treated as a free parameter, while the systematic effect of $B_\text{src}$ is already negligible.

Furthermore, in all analyses the parameters $\mu_{\text{N}_2}$ and $A_{\text{N}_2}$ are treated as nuisance parameters incorporating experimental energy shifts and rate scaling. They are fit with pixel-wise resolution, along with $B_\text{ana}$ and $\mathcal{R}_\text{bg}$.

The parameters $R_{\text{N}_3,\text{N}_2}$ and $\Delta \mu_{\text{N}_3,\text{N}_2}$ are intrinsic to the $\kr$ spectrum and do not depend on the measurement campaign. They were determined from the KrM5 \SI{75}{\percent} measurement to be \num{1.530+-0.013} and \SI{0.665+-0.002}{eV}, respectively (statistical uncertainties only), in agreement with \cite{Venos2018}. The achieved sensitivity allows them to be fixed in all other analyses, which improves the fit convergence.

The $\sigma^2$ fit parameter is constrained to the positive region. The following contributions $\sigma_\mathrm{add}^2$ additional to the source-potential broadening $\sigma_0^2$ are subtracted from $\sigma^2$ after the fit: an $\mathcal{O}(\SI{e-4}{eV^2})$ intra-pixel variance of electric potential in the analyzing plane $\sigma_\mathrm{AP}^2$ obtained from simulations, broadening $\sigma_\mathrm{HV}^2$ coming from instabilities related to the high-vol\-tage system \cite{Rodenbeck_2022} quantified by measurements, and the squared Doppler broadening $\sigma_\text{Doppler}^2$ \cite{Kleesiek2019} caused by the thermal motion of the $\kr$ atoms (cf. table~\ref{tab:HV_broadenings}). It is assumed that thereby all additional broadening contributions are covered and the bare source-potential broadening $\sigma_0^2$ is obtained. If there are further contributions, it would mean that a more conservative limit on $\Delta_{10}$ is set, as discussed in section~\ref{sec:SourceObservables}.

\subsubsection{High-statistics measurement at \SI{75}{\percent} and \SI{40}{\percent} column density}
\label{sec:N23_75percent}

The source-potential broadening both at \SI{75}{\percent} and \SI{40}{\percent} of nominal tritium column density was measured first in the KrM5 campaign, and again in the KrM9 campaign. The analysis of both measurement campaigns is performed in the same way.

In a first step, the temporal stability of the source potential over the duration of the longer, \SI{75}{\percent} column density measurement was confirmed by analyzing the line positions $\mu_{\text{N}_2}$ of the individual scans. Accordingly, all scans within one campaign are combined, allowing for a pixel-resolved analysis. 
For the measurement at \SI{75}{\percent} column density the same 14 patches as in the neutrino-mass analysis are used \cite{KMN1-5-paper}, consisting of 9 pixels each.
For every FPD patch one fit is performed. $\sigma^2$ as well as $R_{\text{N}_3,\text{N}_2}$ and $\Delta \mu_{\text{N}_3,\text{N}_2}$ are common parameters among the pixels in a patch.
Given the high statistics of the measurement at \SI{40}{\percent} column density, an individual analysis for each pixel is performed.

\subsubsection{Measurement at non-optimal rear-wall voltages}

To investigate the dependence of the source-potential broadening on the rear-wall voltage, measurements at \SI{40}{\percent} and \SI{75}{\percent} nominal tritium column density were performed in KrM5. The rear-wall voltage was varied between \SI{-0.6}{V} and \SI{0.2}{V} and all scans for each rear-wall voltage set-point and tritium column density are combined.
Pixel-resolved fits are performed, and the fit results are averaged over the whole detector.

\section{Results and impact for the KATRIN experiment}
\label{sec:results}

\begin{figure}[h]
	\centering
		\includegraphics[width=0.49\textwidth]{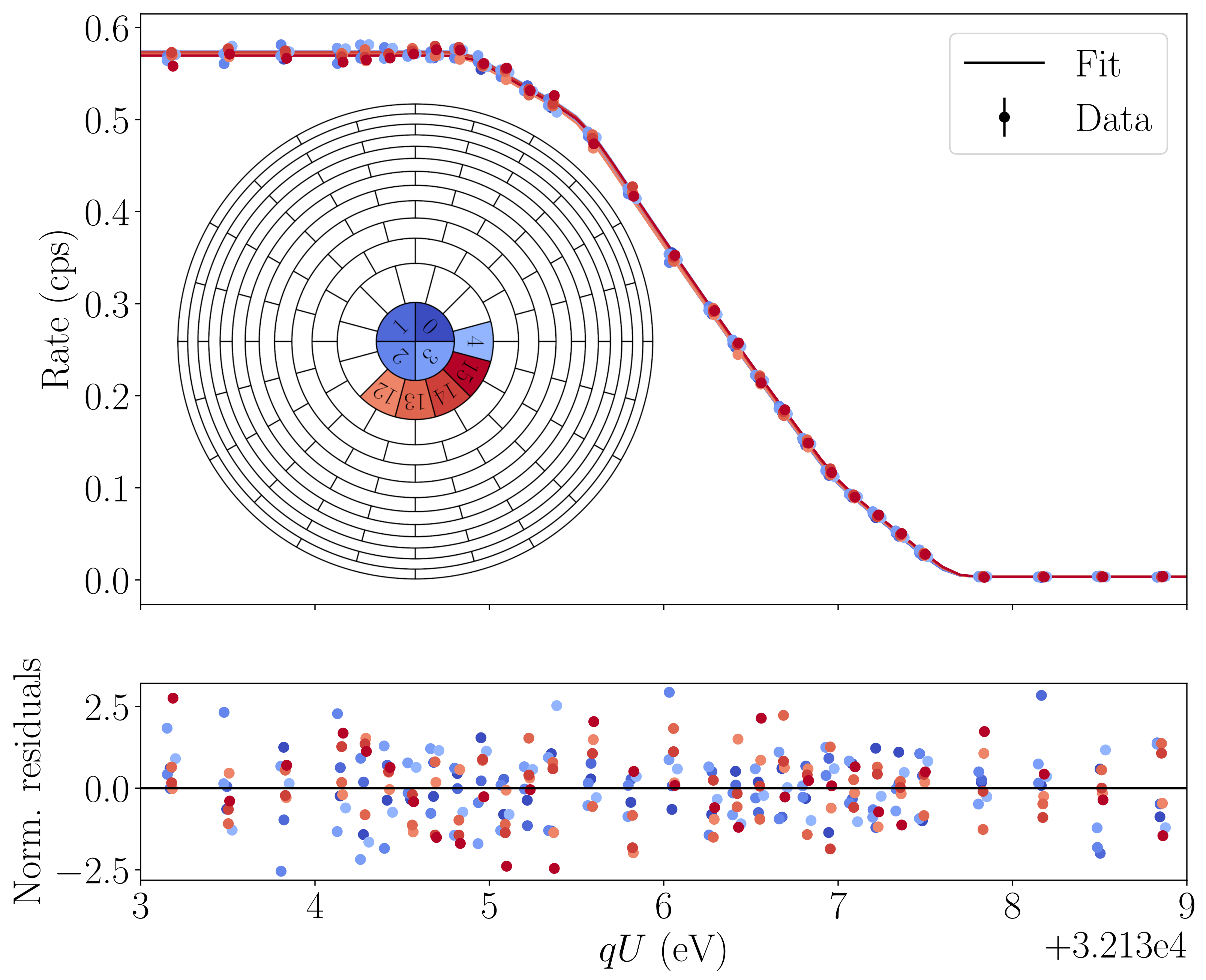}
	\caption{Spectral fit of patch 0 of the \SI{75}{\percent} column-density measurement taken during KrM5. The recorded detector count rate, fit and normalized residuals are shown over the retarding potential set points $qU$. The numbers and positions of the pixels of patch 0 are shown in the FPD-wafer schematic. }
	\label{fig:Spectrum}
\end{figure}
Following the aforementioned analysis procedure, we present the results of the $\kr$ measurement campaigns KrM5 and KrM9. Figure~\ref{fig:Spectrum} shows an example fit of one of the KrM5 spectra.

First the radial dependency of the line broadening measured at the optimal rear-wall voltage is shown and the mean values are discussed. Furthermore, the impact of a non-op\-ti\-mal rear-wall voltage set-point is investigated and finally the implications for the neutrino-mass analysis are presented.

\subsection{Source-potential broadening results of the high-statistics measurements}

\begin{figure}[h]
	\centering
			\includegraphics[width=0.49\textwidth]{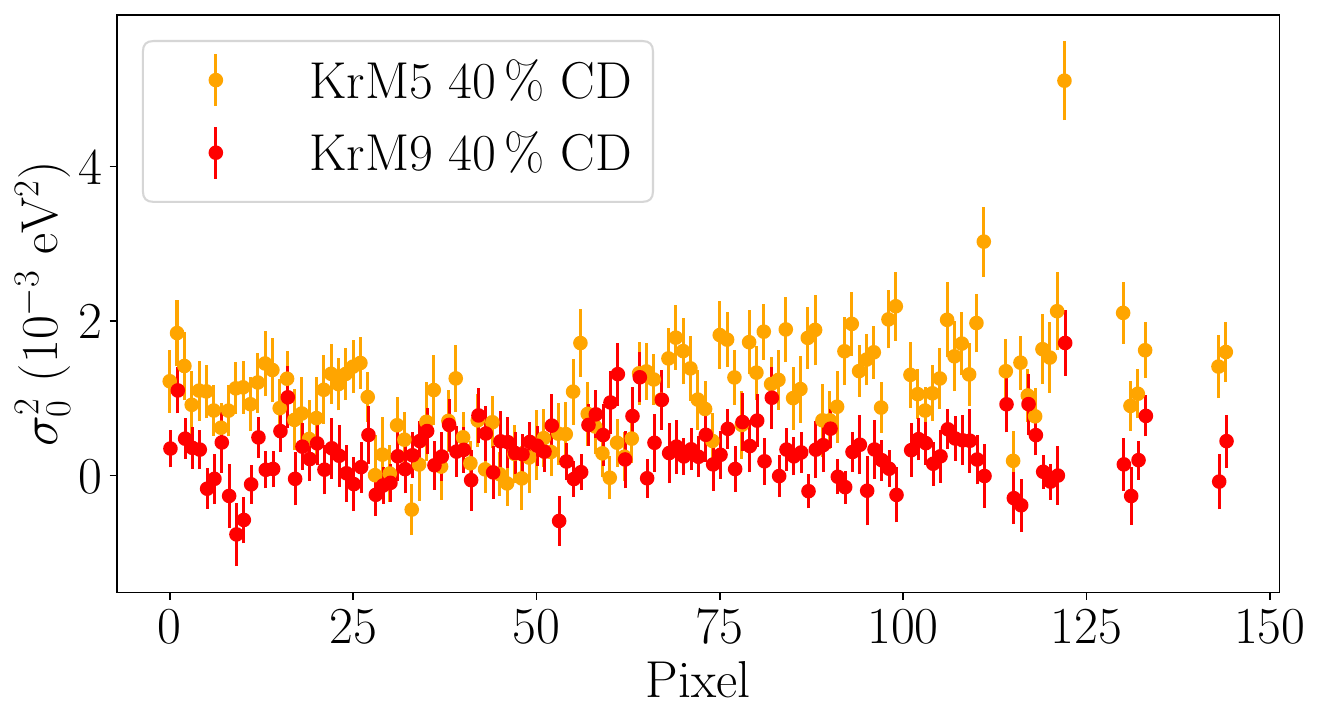}
			\includegraphics[width=0.49\textwidth]{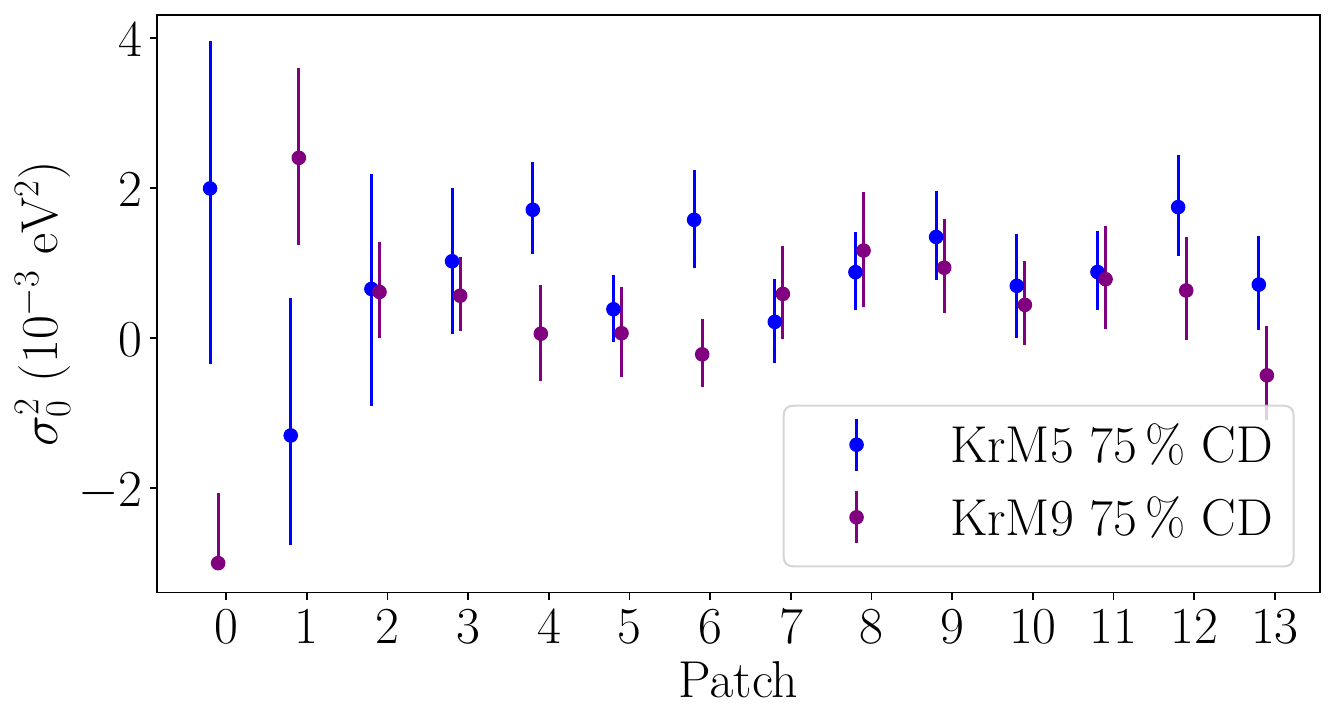}
	\caption{Squared source-potential broadening $\sigma_0^2$ for the \SI{40}{\percent} and \SI{75}{\percent} column-density measurements over radial FPD resolution, comparing the measurement campaigns KrM5 and KrM9. From the bare fit broadening $\sigma^2$, which is non-negative, additional broadening contributions $\sigma^2_\mathrm{add}$ have been subtracted. Their uncertainties are not shown, since they are fully correlated over all data points in a data set. Two noteworthy features are visible: A dip in the broadening between pixel 25 to 50 and an oscillating pattern in the KrM5 measurement at \SI{40}{\percent} column density (yellow), and a small broadening in patch 0 of the KrM9 data set at \SI{75}{\percent} column density (purple).}
	\label{fig:BroadeningSquared}
\end{figure}

In figure \ref{fig:BroadeningSquared}, the radially dependent source-potential broadening result from the high-statistics measurements is shown. We observe two features: Firstly, we see an oscillating pattern of $\sigma_0^2$ that increases for outer pixels and has an overall drop in the range between pixel 25 and 50 in the KrM5 measurement at \SI{40}{\percent} column density. Secondly, there is a quite small broadening of patch 0 in the KrM9 measurement at \SI{75}{\percent} column density. Despite thorough investigations of the analysis strategy, data quality, and experimental conditions, no origin of these features was found. The feature in the KrM5 \SI{40}{\percent} data set is assumed to be a physical effect of the source. This is the only dataset that showed on average increased $\chi^2/\text{ndf}=1.38$ for 35 degrees of freedom (ndf) in the fits, which was accounted for by increasing the uncertainties by $\sqrt{\chi^2/\text{ndf}}$. The feature in KrM9 is assumed to be a statistical fluctuation. Beyond that, good homogeneity is observed. Thus, we construct an average value of the broadening over the radius; for the \SI{40}{\percent} measurements we use a constant fit, while in the \SI{75}{\percent} measurements we perform a likelihood profiling over the broadening to correctly consider the asymmetrical uncertainties of patch 0. The final values are shown in table \ref{tab:Final_broadenings}.

\begin{table}[h]
    \centering
    \caption{Mean squared source-potential broadenings obtained from the N-32 measurements at the optimal rear-wall voltage. The values are uniform averages of the measurements and include the corrections $\sigma^2_\mathrm{add}$ discussed in section~\ref{sec:NAnalysis}, which dominate the uncertainties.}
    \begin{tabular*}{\columnwidth}{@{\extracolsep{\fill}}lcc@{}}
    \toprule
        &\SI{40}{\percent} CD & \SI{75}{\percent} CD  \\
        \midrule
       $\sigma_0^2$ (\SI{e-3}{eV^2}) in KrM5 & \num{1.0+-0.2} & \num{1.0+-0.2}\\
       $\sigma_0^2$ (\SI{e-3}{eV^2}) in KrM9 & \num{0.3+-0.1} & \num{0.4+-0.2}\\
        \bottomrule
        \end{tabular*}
    \label{tab:Final_broadenings}
\end{table}

The broadening values for the measurements at \SI{40}{\percent} and \SI{75}{\percent} column density of the same campaign agree within uncertainties. This implies that the source potential is not dominated by the density of the charges within the source. However, the values differ between the two campaigns, indicating that the source potential is more influenced by the work functions of the source walls.

One effect producing a source-potential broadening could be an inhomogeneity of the work functions, which may change over time, e.g. because of accumulation of tritium on the surfaces.
Inhomogeneous work functions could be responsible for the features in figure~\ref{fig:BroadeningSquared}, as well as the difference in the mean values between the krypton campaigns.

The surface conditions can change not only between, but also within neutrino-mass measurement campaigns. Thus, drifting work functions can lead to a non-optimal rear-wall set point, and the effect on the source-potential broadening is discussed in the following section.

\subsection{Rear-wall voltage dependency of the source-potential broadening}

\begin{figure}[]
	\centering
		\centering
		\includegraphics[width=0.49\textwidth]{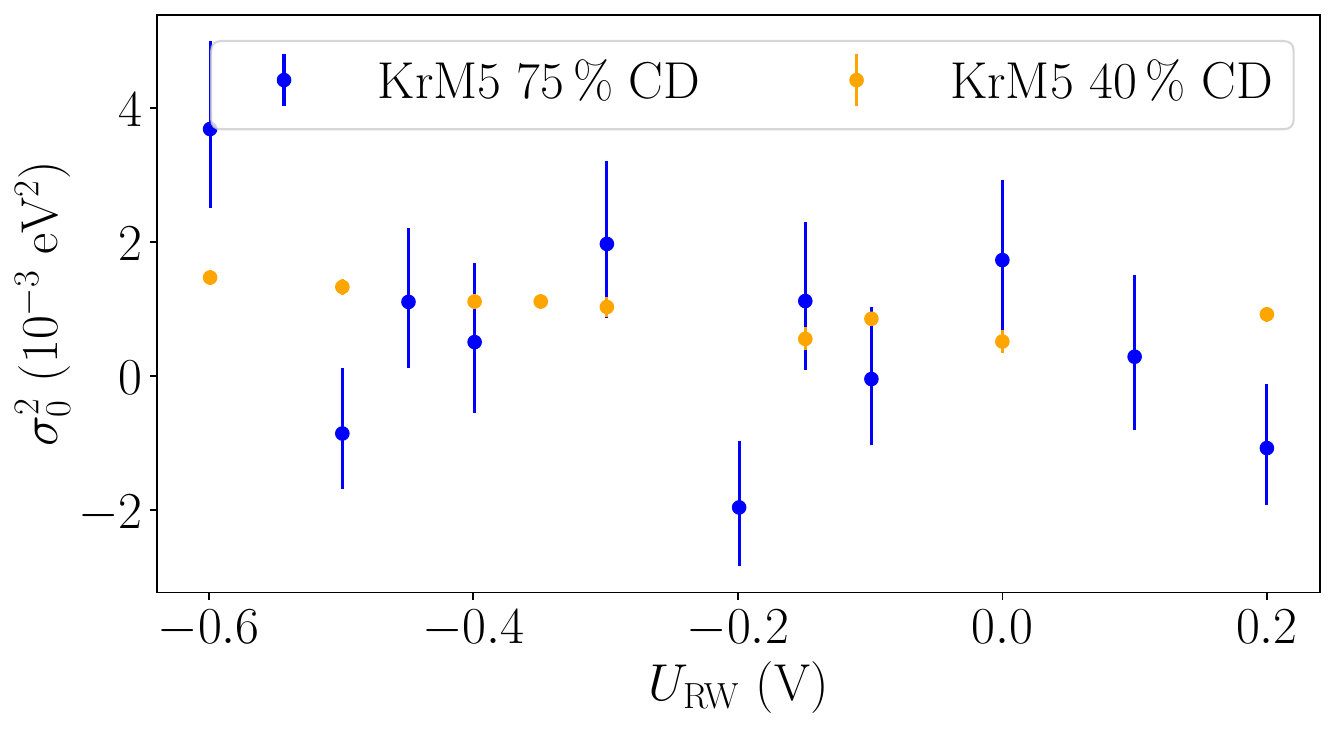}
	\caption{Source-potential broadening $\sigma_0^2$ as a function of the rear-wall voltage, measured at \SI{75}{\percent} and \SI{40}{\percent} column density. From the bare fit broadening $\sigma^2$ of the N$_{23}$-32 lines, which is constrained to the non-negative parameter region, contributions $\sigma^2_\mathrm{add}$ have been subtracted. The error bars contain statistical and high-voltage-related uncertainty, which is uncorrelated between the data points. A slight slope of $\SI{-2.0+-0.6e-3}{eV^2/V}$ within an interval of \SI{0.4}{V} around the optimal rear-wall voltage of \SI{-0.3}{V} is observed for the measurement at \SI{40}{\percent} column density. This dependency is not visible in the measurement at \SI{75}{\percent} column density within the margin of errors. 
    }
	\label{fig:Broadenings_vs_RW_reduced_withDopp}
\end{figure}

In figure \ref{fig:Broadenings_vs_RW_reduced_withDopp}, the source-potential broadening as a function of the rear-wall voltage at \SI{40}{\percent} column density and \SI{75}{\percent} column density is shown. Within their uncertainties both measurements are consistent with each other, and the broadening at the optimum voltage is also consistent with the high-statistics measurement. In the high-statistics \SI{40}{\percent} measurement, a significant dependence of the broadening on the rear-wall voltage can be observed.
Based on measurements of the rear wall current \cite{Friedel2020} we expect a maximum deviation of $\left|U_\text{RW,opt}-U_\text{RW,set}\right|=\SI{0.1}{V}$ of the optimal rear-wall voltage from the rear-wall voltage set point in neutrino-mass measurements. A linear function fit to the \SI{40}{\percent} column-density values is applicable within an interval of \SI{0.4}{V} around the optimal rear-wall voltage of \SI{-0.3}{V}, yielding a slope of $\frac{d \sigma_0^2}{d U_\text{RW}}=\SI{-2.0+-0.6e-3}{eV^2/V}$.
Based on this, an additional systematic uncertainty on the broadening is taken into account, following
\begin{equation}
\label{eq:sys_rw}
\delta\sigma^2_\text{opt,RW}=\left|U_\text{RW,opt}-U_\text{RW,set}\right|\cdot \frac{d \sigma_0^2}{d U_\text{RW}} = \SI{0.2e-3}{eV^2}.
\end{equation}

\subsection{Derived source-potential observables for the neutrino-mass analysis}

As shown in the previous sections, the source-potential broadening obtained differs between the $\kr$ measurement campaigns. This could be an effect of the surface conditions of the source. The KNM1-5 analysis is based only on the measured source-potential broadening of KrM5, as this leads to a more conservative estimation of the systematic effect on the neutrino mass. It is important to acknowledge that subsequent to the KrM5 measurement, two rear wall cleaning procedures employing UV/ozone treatments were conducted, leading to a significant modification of the surface conditions. 

The neutrino-mass measurement campaigns of KNM1-5 were taken at four different tritium column densities and at different temperatures. Since no column-density dependence of the source-potential broadening was observed, the same mean broadening for all neutrino-mass measurement campaigns was used in the KNM1-5 analysis \cite{KMN1-5-paper}. To account for the differing temperatures of KNM1\&2 compared to the $\kr$ measurements, the uncertainty of the broadening is set equal to its mean to be conservative. Even with this conservative approach, the source potential is not among the leading systematic effects in KNM1\&2.

The inputs for the neutrino-mass campaigns are shown in table \ref{tab:FinalInputs}. The upper limit of the effective asymmetry parameter $\Delta_\text{P}$ is calculated using the principle laid out in equation \ref{equ:delta10}, with the individual source-potential broadenings for each campaign and $\kappa$ coefficients fitting to the corresponding column density.

\begin{table*}[]
    \centering
    \caption{Source-potential inputs for the neutrino-mass analysis for different campaigns (Naming of the campaigns in accordance with \cite{KMN1-5-paper}). The source-potential broadenings are derived from the KrM5 values from table~\ref{tab:Final_broadenings}, accounting for an additional systematic uncertainty due to possible deviation from the optimal rear-wall voltage during neutrino-mass measurements, and an extrapolation for KNM1\&2. The values of the effective parameter $\Delta_\text{P}$ are calculated following section~\ref{sec:SourceObservables}.}
    \begin{tabular*}{\textwidth}{@{\extracolsep{\fill}}lclcc@{}}
    \toprule
          Campaign &Configuration & Data set  & $\sigma_0^2$ (\SI{e-3}{eV^2}) & $\Delta_\text{P}$ (meV) \\
    \midrule
    KNM1 &   \SI{20}{\percent} CD, \SI{30}{\kelvin} & Extrapolation from KrM5  & $\num{1\pm 1}$ & $\num{0\pm20}$\\
     KNM2 &     \SI{84}{\percent} CD, \SI{30}{\kelvin} & Extrapolation from KrM5  &  $\num{1\pm 1}$ & $\num{0\pm24}$  \\  
      KNM3-SAP &   \SI{40}{\percent} CD, \SI{80}{\kelvin} & KrM5, \SI{40}{\percent} CD  & $\num{1.0\pm 0.3}$ & $\num{0\pm20}$\\
     KNM3-NAP, KNM4, KNM5 &     \SI{75}{\percent} CD, \SI{80}{\kelvin} & KrM5, \SI{75}{\percent} CD & $\num{1.0\pm 0.3}$ & $\num{0\pm22}$\\
    \bottomrule
    \end{tabular*}
    \label{tab:FinalInputs}
\end{table*}
The values shown in table \ref{tab:FinalInputs} differ slightly compared to the values used in the most recent neutrino-mass analysis \cite{KMN1-5-paper}. The values shown here present our improved knowledge, including the effect of the angular dependent detection efficiency and the additional uncertainty due to the non-optimal rear-wall voltage set point. The changes of the source-potential broadening are within one third of the now slightly larger uncertainty. The dominating input, the limit on the asymmetry parameter, changes insignificantly by \SI{1}{mV}. The related change of the estimated squared neutrino mass on the order of \SI{e-3}{eV^2} is negligible.

\section{Conclusions and outlook}

The understanding of systematic effects that modify the electron spectrum of $\upbeta$-electrons is essential to achieve a high sensitivity on the neutrino mass with the KATRIN experiment. Of key importance is the knowledge of the inhomogeneity of the electric potential in the tritium source in which the electrons are generated.
$\upbeta$-decay and subsequent secondary ionization, dissociation, and recombination processes lead to the formation of a cold plasma, whose electric potential distribution is influenced by external fields, the gas density and flow, the work functions of the source, and the voltage applied to the rear wall of the experiment. Since numerical approaches for the quantification of these effects proved to be insufficient, in-situ measurements were adopted to determine the source-potential inhomogeneity. The latter is quantified by two observables: a Gaussian broadening $\sigma_0$, measuring the overall inhomogeneity, and a parameter $\Delta_\text{P}$, measuring the asymmetry of the potential with respect to the central injection point of the source.
Precision spectroscopy of conversion electrons from the meta-stable $\kr$ co-cir\-cu\-la\-ted with the tritium gas in the KATRIN source has been refined over the last years with regard to the theoretical understanding of the concept \cite{Moritz-paper,Machatschek2021} and the technical implementation \cite{Marsteller2022}. Subsequent optimizations of the performing of effective measurement campaigns with a high activity $^{83}$Rb/$\kr$ source, and the accurate inference of the source-potential observables from the data are reported in this work.

The source-potential broadening for different gas densities was measured in two high-statistics measurement campaigns, KrM5 in 2021 and KrM9 in 2023. Within the campaigns the broadening does not show a dependency on the gas density, while it differs by up to 3 sigma between the campaigns. To quantify the effect of the source-potential inhomogeneity on the neutrino-mass analysis we use the more conservative broadening estimate. While the data recorded in the KrM5 and KrM9 campaigns also allows the determination of the $\Delta_\text{P}$ parameter, currently it is dominated by systematic effects related to the modeling of the intrinsic $\kr$ spectrum. Instead, we use a phenomenological approach to constrain $\Delta_\text{P}$ by the measured broadening in the most conservative way.

With the estimates of the source-potential inhomogeneity derived in this work one finds a neutrino-mass systematic effect on the order of \SI{2e-2}{eV^2}, which is sufficient for the intermediate neutrino-mass result of KATRIN \cite{KMN1-5-paper}, but exceeds the 0.0075-eV$^2$ systematic budget for the final KATRIN result \cite{KATRINdesign}. By understanding the systematic effects in the $\Delta_\text{P}$ analysis, it is possible to reduce the source-potential related neutrino-mass systematic by a factor of up to 3 with the existing $\kr$ data. This would be sufficient for the final KATRIN analysis.

However, the difference observed between the two krypton campaigns led us to question the long term reliability of the source-potential estimates. Currently, work-function drifts are being considered as the main origin of this effect. In 2025, a concluding set of $\kr$ measurements will help us to understand the underlying physical processes and their effect on the observables, allowing reliable estimates for the final KATRIN analysis.

\begin{acknowledgements}
We acknowledge the support of Helmholtz Association (HGF), Ministry for Education and Research BMBF (05A23PMA, 05A23PX2, 05A23VK2, and 05A23WO6), the doctoral school KSETA at KIT, Helmholtz Initiative and Networking Fund (grant agreement W2/W3-118), Max Planck Research Group (MaxPlanck@TUM), and Deutsche Forschungsgemeinschaft DFG (GRK 2149 and SFB-1258 and under Germany's Excellence Strategy EXC 2094 – 390783311) in Germany; Ministry of Education, Youth and Sport (CANAM-LM2015056) in the Czech Republic; Istituto Nazionale di Fisica Nucleare (INFN) in Italy; the National Science, Research and Innovation Fund via the Program Management Unit for Human Resources \& Institutional Development, Research and Innovation (grant B39G670017) in Thailand; and the Department of Energy through Awards DE-FG02-97ER41020, DE-FG02-94ER40818, DE-SC0004036, DE-FG02-97ER41033, DE-FG02-97ER41041,  {DE-SC0011091 and DE-SC0019304 and the Federal Prime Agreement DE-AC02-05CH11231} in the United States. This project has received funding from the European Research Council (ERC) under the European Union Horizon 2020 research and innovation programme (grant agreement No. 852845). We thank the computing support at the Institute for Astroparticle Physics at Karlsruhe Institute of Technology, Max Planck Computing and Data Facility (MPCDF), and the National Energy Research Scientific Computing Center (NERSC) at Lawrence Berkeley National Laboratory.
\end{acknowledgements}

\bibliographystyle{unsrtnat}
\bibliography{krypton-paper-references}

\end{document}